# $h\nu^2$-concept breaks the photon-count limit of RIXS instrumentation

*Kejin Zhou[1]\*, Satoshi Matsuyama[2] & Vladimir N. Strocov[3]\**

[1] Diamond Light Source, Harwell Campus, Didcot OX11 0DE, United Kingdom
[2] Department of Precision Science and Technology, Graduate School of Engineering, Osaka University, 2-1 Yamada-oka, Suita, Osaka, Japan
[3] Swiss Light Source, Paul Scherrer Institute, 5232 Villigen-PSI, Switzerland
\* Correspondence e-mails: kejin.zhou@diamond.ac.uk; vladimir.strocov@psi.ch

## Abstract

Upon progressive refinement of energy resolution, the conventional RIXS instrumentation reaches the limit where the bandwidth of incident photons becomes insufficient to deliver an acceptable photon-count rate. We show that the RIXS spectra as a function of energy loss are essentially invariant to their integration over incident energies within the core-hole lifetime. This fact permits the RIXS instrumentation based on the *$h\nu^2$*-concept to utilize incident synchrotron radiation over the whole core-hole lifetime window without any compromise on the energy-loss resolution, thereby breaking the photon-count limit.

# Introduction

Synchrotron-radiation based resonant inelastic X-ray scattering (RIXS) is one of the most advanced spectroscopic techniques giving access to the whole spectrum of charge-neutral excitations in condensed and soft matter, ranging from the charge transfer and orbital excitations on its high-energy end of a few eV, to magnon, phonon and many exotic elementary excitations on its low-energy end extending ultimately down to a few meV [1-7]. High energy resolution is a prerequisite of the RIXS experiment in pursuit of the progressively refining energy scale of these excitations.

During the last decade, the RIXS instrumentation has seen enormous progress, with its state-of-art resolving power now approaching 50,000. The increase comes on both the incident photon side, defined by the beamline energy resolution, and on the outgoing photon side defined by the spectrometer energy resolution. At a certain stage of the resolution improvement, however, the incident photon energy bandwidth reduces so much that the total number of photons delivered to the sample becomes insufficient for the accumulation of RIXS signal within any reasonable acquisition time. We call this situation the *photon-count limit* of RIXS instrumentation. It cannot be circumvented on the route of the conventional RIXS instrumentation, where the resolution improvement inevitably reduces the utilized bandwidth of incident X-ray photons.

The way to break through the photon-count limit can be envisaged by bringing together insights into the fundamental physics of the RIXS process and novel instrumental concepts. On the physics side, the ability of RIXS to detect elementary excitations owes to the intermediate core-hole state. For instance, the spin-flip excitations at the *L*-edge in many transition metal oxides are allowed thanks to the strong spin-orbit (LS) coupling of the 2*p* core-hole state. Importantly, the intermediate state is naturally broadened due to the core-hole lifetime. Although the spectrum of the elementary excitations observed in RIXS critically depends on the incident energy involving different intermediate states, one can nevertheless expect that this spectrum stays constant as long as the incident energy is kept within the lifetime broadening of the selected intermediate state.

On the instrumental side, a way to utilize an extended bandwidth of incident-photon energy ($h\nu_{in}$) without any compromise on resolution in outgoing-photon energy ($h\nu_{out}$) can be realized with the recently suggested $h\nu^2$-concept of the RIXS spectrometer [8]. The line focus of light from the focal plane of the beamline monochromator, dispersed in $h\nu_{in}$ in the vertical direction, is imaged by refocusing optics onto the sample. The spectrometer uses a focusing mirror in its vertical (imaging) plane to refocus the image from the sample onto the position-sensitive detector, resolving the scattered light in $h\nu_{in}$. At the same time, a cylinder VLS grating operating in the horizontal (dispersive) plane of the spectrometer disperses and focuses the scattered light onto the detector, resolving it in $h\nu_{out}$. In this way the RIXS intensity is acquired as a two-dimensional (2D) image in the $h\nu_{in}$- and $h\nu_{out}$-coordinates detected simultaneously [8]. Obviously, this concept implies sample homogeneity within the line focus on the sample, which can though be squeezed by the refocusing optics to ~100 μm and less. Later, Warwick *et al* realized that the main factor limiting the $h\nu_{in}$-acceptance and thus efficiency of this concept is the spatial extension of the light footprint on the sample along the $h\nu_{in}$-direction, and suggested to solve this problem by replacing the focusing mirror in the imaging plane of the spectrometer with a Wolter-type imaging optics [9]. Furthermore, the focal-plane inclination at the sample has been eliminated with an elliptical pre-mirror of the monochromator. The first project to realize this scheme, the double-dispersion QERLIN spectrometer at the ALS, is presently nearing its completion [10]. Remarkably, a dramatic increase of the overall scattered X-ray intensity detected with such an imaging-$h\nu^2$ spectrometer of about two orders of magnitude compared to the conventional state-of-art RIXS spectrometers comes without any compromise on the energy-loss resolution.

Here, we demonstrate fundamental scientific grounds allowing a breakthrough of the photon-count limit of RIXS instrumentation on its way of progressively refining energy resolution. We report ultra-high resolution RIXS experiments on a prototypical binary oxide CoO across the Co $L_3$-edge, and demonstrate that the RIXS spectral line profile on the energy-loss scale stays invariant of $h\nu_{in}$ as long as kept within the core-hole lifetime broadening. Furthermore, we analyze the RIXS intensity distribution in the ($h\nu_{in}$,$h\nu_{out}$)-coordinates generated by the imaging-$h\nu^2$ spectrometer and demonstrate the way of its rendering into the RIXS spectra over the whole $h\nu_{in}$-window of the core-hole lifetime, increasing the efficiency of the RIXS experiment by one-two orders of magnitude without compromising its resolution on the energy-loss scale.

# RIXS spectra across the absorption line

Ultra-high resolution experiments to explore the sensitivity of RIXS spectral structures as a function of $h\nu_{out}$ to variation of $h\nu_{in}$ were performed at the I21-RIXS beamline at Diamond Light Source, United Kingdom [11]. Single crystal CoO was selected for the study focused on the low-energy excitations at the Co $L_3$-edge. The sample was aligned with the surface normal (100) lying in the horizontal scattering plane. Co $L_3$-edge XAS was collected in the fluorescence yield. For RIXS measurements, linear *σ*-polarization was used. The total energy resolution d*E* at the full width at half maximum (FWHM) was about 33 meV at the Co $L_3$-edge. The grazing-incidence and scattering angles *θ* and 2*θ* were fixed to 20° and 154°, respectively. All measurements were done at a temperature of 13 K.

Fig.1(a) displays the Co $L_3$-edge XAS. The fine structure is dominated by the multiplet effects of the intermediate core-hole states of the $Co^{2+}$ ion within its crystal-field symmetry. Two $h\nu_{in}$ regions were chosen (red and blue highlights) to excite RIXS in a step of 0.03 eV. The low-energy RIXS spectra in these regions, presented in Fig. 1(b) and (c), were obtained by summation over $h\nu_{in}$-intervals of different energy windows. For example, the spectrum denoted by ±0.00 eV means it was acquired at the central incident energies $h\nu_{in}$ = 777.14 eV and 778.99 eV of the two regions, while the spectrum denoted by ±0.15 eV is the average of the RIXS spectra excited with $h\nu_{in}$ from -0.15 eV to +0.15 eV relative to the central incident energies. A three-peak structure is clearly present in all RIXS spectra. The shoulder peak at zero energy results from the quasi-elastic scattering from the sample, and the second and third peaks at ~60 meV and ~120 meV are arguably dominated by the one-magnon and two-magnon excitations, respectively [12].

It is noticeable that the line shape of the averaged RIXS spectra deviates more and more from that of the central RIXS spectrum as the $h\nu_{in}$-window widens. To quantify our analysis, we define $P = (I_{ave}(E) - I_0(E)) / I_0(E)$ as a measure of the amount of deviation, in which $I_{ave}(E)$ represents the RIXS spectra averaged over the $h\nu_{in}$-window and $I_0(E)$ stands for the RIXS spectrum excited by the central $h\nu_{in}$. The value of $P$, plotted in Fig.1(d) and (e), shows a monotonic increase as the $h\nu_{in}$-window widens to ±250 meV. We can define a threshold, below which the RIXS spectra are regarded as invariant, as $P = 10\%$. This corresponds to an $h\nu_{in}$-integration window of about ±0.24 eV which is roughly the FWHM of the Co $L_3$ core-hole lifetime broadening 0.43 eV [13].

In this way, we have shown that the RIXS spectra can be integrated in an $h\nu_{in}$-window of the core-hole lifetime while keeping the whole spectroscopic information. We note that our case of magnon excitations can be considered as one of the most critical on the $h\nu_{in}$-bandwidth, and high energy-loss excitations, for example, the orbital excitations, are typically less sensitive to it.

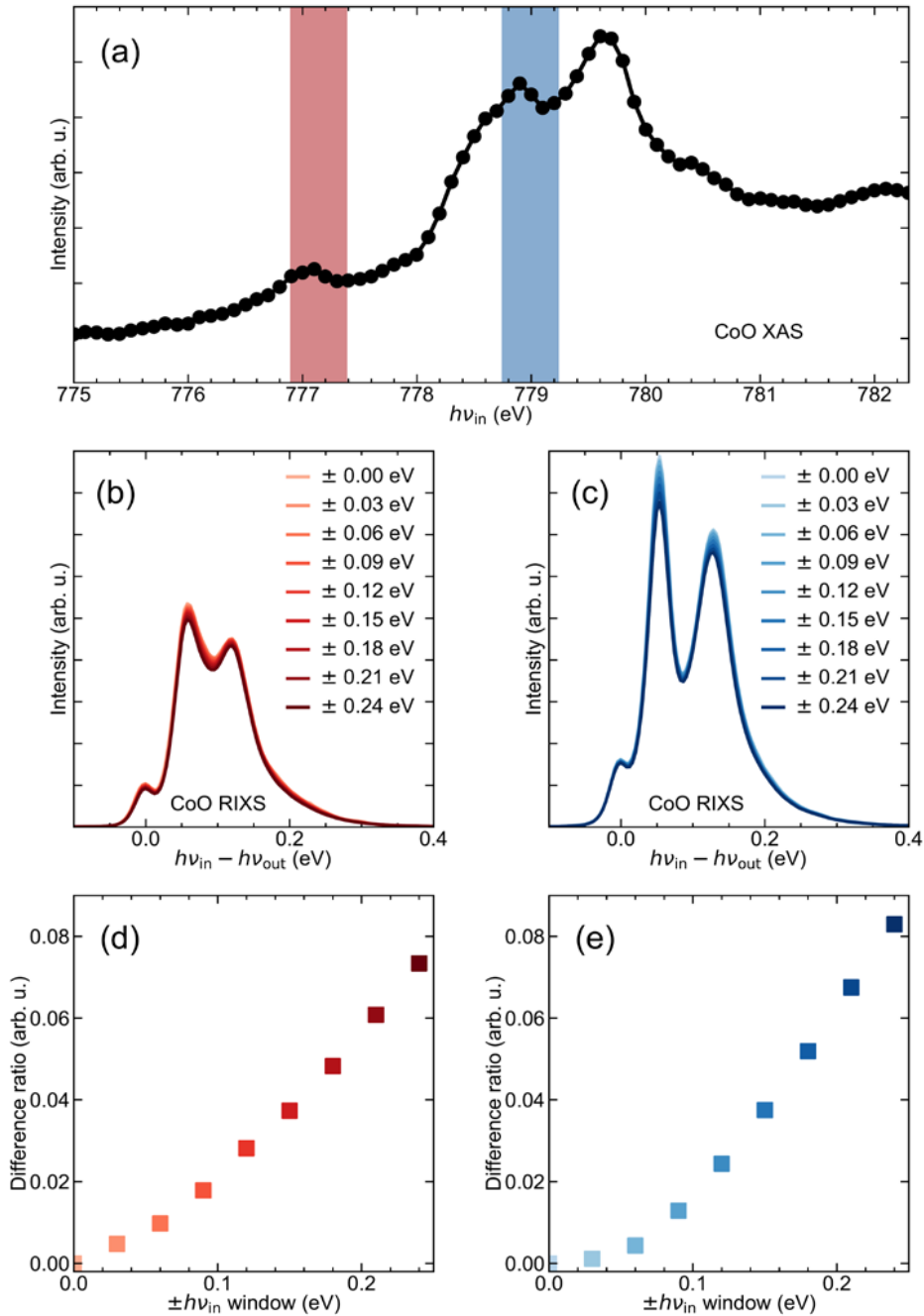

Fig. 1(a) The Co $L_3$-edge XAS spectrum collected using fluorescence yield. Red and the blue bars mark the energy windows where the RIXS spectra were acquired. (b) and (c) Low-energy excitations in CoO as a function of the $h\nu_{in}$-integration window for the region marked in red and blue in (a), respectively. (d) and (e) are the difference ratio $P$ between the integrated and central RIXS spectra as a function of the $h\nu_{in}$-window (see the definition in the main text).

# Detection efficiency of the $h\nu^2$-concept

We now demonstrate how the $h\nu^2$-concept translates the invariance of the RIXS spectra across the absorption line into a dramatic detection-efficiency increase. Fig. 2 schematizes a XAS spectral peak and corresponding RIXS intensity map, where the elastic peak stretching along the $h\nu_{in}=h\nu_{out}$ line is tracked by two Raman satellites at constant energy loss $h\nu_{in}$-$h\nu_{out}$. The sought-for spectrum of RIXS intensity as a function of energy loss, delivered by the conventional spectrometer, is intrinsically integrated over $h\nu_{in}$ within the bandwidth $\delta E$ determined by the required combined beamline + spectrometer energy resolution $dE$ as $\delta E = dE/\sqrt{2}$ (in the ideal case when the two contributions are balanced). In contrast, the full 2D map acquired with the imaging-$h\nu^2$ spectrometer allows us to obtain the RIXS spectrum by integrating along the $h\nu_{in}$-$h\nu_{out}$ direction within a much larger interval $\Delta E$ determined by the XAS peakwidth. In this way, the $h\nu^2$-concept allows full utilization of a broad $h\nu_{in}$ bandwidth, realistically up to two orders of magnitude compared to the energy-loss resolution, which dramatically increases the detection efficiency without any sacrifice on the latter. It should be noted that the use of imaging optics in the refocusing and, as suggested by Warwick at al [9], spectrometer stages of the optical scheme ensures the invariance of energy resolution over the whole intercepted $h\nu_{in}$ range. We note that another concept that promises accepting a large $h\nu_{in}$ bandwidth without sacrificing the energy-loss resolution is the active-grating monochromator (AGM) and spectrometer (AGS) implemented at the Taiwan Photon Source [14]. Such an instrument inherently integrates the RIXS intensity along the $h\nu_{in}$-$h\nu_{out}$ direction to deliver essentially a 1D spectrum, whereas the $h\nu^2$-concept delivers the full 2D map allowing integration either along constant $h\nu_{in}$-$h\nu_{out}$ to keep energy resolution of the Raman peaks, or along constant $h\nu_{out}$ to keep that of the fluorescence peaks.

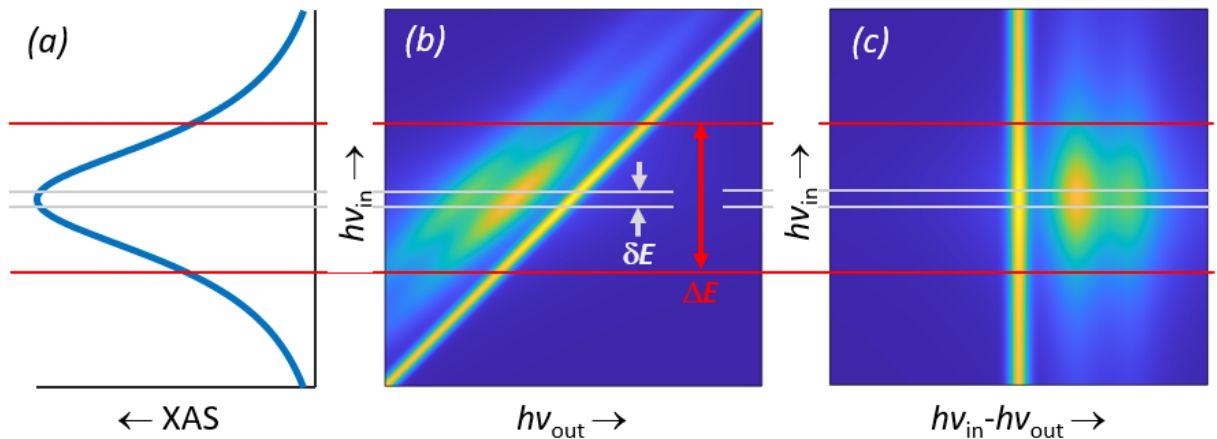

Fig. 2. Schematic XAS spectrum (*a*) and the corresponding RIXS intensity map in the ($h\nu_{in}$,$h\nu_{out}$)-coordinates (*b*) and rendered into the ($h\nu_{in}$,$h\nu_{in}$-$h\nu_{out}$)-coordinates (*c*). Whereas the conventional RIXS spectrometer utilizes incident photons in a narrow bandwidth $\delta E$, the $h\nu^2$ one utilizes a much larger bandwidth $\Delta E$ of the XAS peakwidth.

Finally, we will assess the detection-efficiency gain delivered by the imaging-$h\nu^2$ spectrometer relative to the conventional RIXS spectrometer adopting the highest-transmission optical scheme consisting of a collecting mirror and a spherical VLS-grating [15, 16]. For the imaging-$h\nu^2$ spectrometer we assume a refocusing-optics stage employing two imaging pairs operating in the horizontal and vertical planes, and for the conventional one of a Kirkpartik-Baetz pair of ellipsoidal mirrors. The gain is defined essentially by the ratio of the FWHM core-level $\Delta E_{CL}$, utilized by the former, to the beamline FWHM resolution, utilized by the latter. This ratio should be further reduced to account for the three additional mirrors in the imaging-$h\nu^2$ scheme compared to the conventional one (two additional reflections in the imaging-optics refocusing stage and one in the imaging stage of spectrometer [9, 10], a factor of ~0.9 at each) and for another factor of ~0.9 describing the average amplitude of the XAS peak within its FWHM relative to the peak value. The results are presented in Fig. 3 as a function of combined d$E$ for three values of $\Delta E_{CL}$. Obviously, the gain is proportional to $\Delta E_{CL}$ and blows up with the resolution refinement. Even for the lowest $\Delta E_{CL}$= 250 meV, the imaging-$h\nu^2$ spectrometer overtakes the conventional one starting already from d$E$ ~ 115 meV, achieves a detection-efficiency advantage of ~4 at the currently-standard d$E$ ~ 30 meV, and boosts this advantage in a singular manner with further refinement of d$E$. These results clearly show the breakthrough of the photon-count limit on the way of progressive energy-resolution refinement. In free speech, it is really astonishing to see how all incident photons within a bandwidth, say, of 1 eV build up a RIXS spectrum with a resolution of 10 meV.

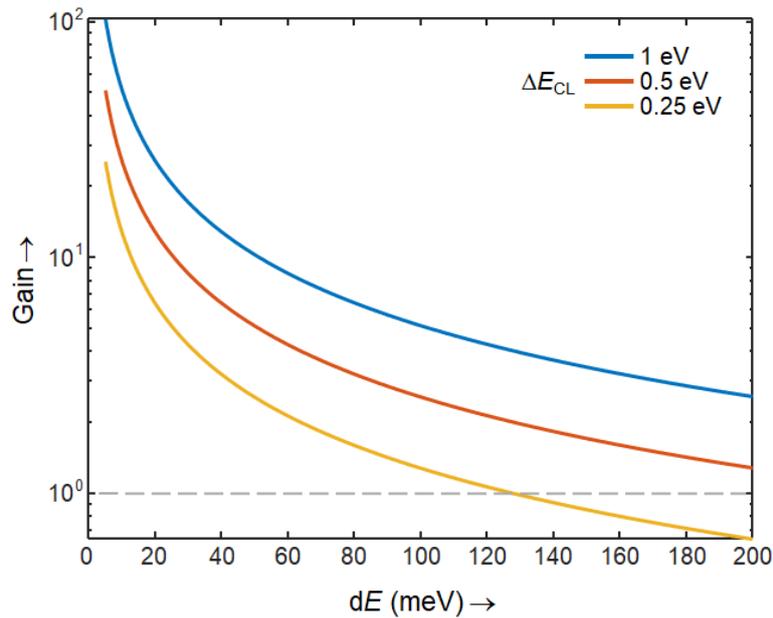

Fig. 3. Intensity gain delivered by the imaging-$h\nu^2$ spectrometer relative to the conventional one as a function of energy-loss resolution for different core-level energy widths.

We will now illustrate these considerations with our experimental example at the Co $L_3$-edge presented above in Fig. 1, where the total resolution d$E$ was ~33 meV at the incident X-ray beam bandwidth $\delta E$ ~ 23 meV. The latter corresponds to an exit slit opening of 10 microns. If the $h\nu^2$-concept is used here, one can have the same total d$E$ while opening the exit slit to 200 microns delivering $\delta E$ ~ 480 meV. In simple terms, one can imagine that twenty horizontal stripes of X-ray beam are laid vertically on the sample, with each stripe delivering a RIXS spectrum identical to the central one. Eventually, one sums up all of them into the total RIXS spectrum without loss of the total d$E$. Thus the efficiency gain achieved by the imaging-$h\nu^2$ spectrometer by using the whole bandwidth of incident X-ray within the core-hole lifetime FWHM, even corrected for the additional mirrors in the optical scheme, is nearly a factor of 15 compared to the conventional RIXS instrumentation.

We note that in many cases such as orbital excitations the RIXS spectra are less sensitive to the $h\nu_{in}$-bandwidth. Furthermore, many experiments focus merely on energy and not the magnitude or lineshape of the RIXS structures. In such cases the $h\nu^2$ concept allows utilization of incident photons through the whole energy extension of the multiplet and spin-orbit structure of the XAS spectrum, allowing much larger intensity gain compared to the core-lifetime window. This situation has a similarity to non-resonant X-ray fluorescence measurements, where one opens up the beamline exit slit to utilize incident photons in a large $h\nu_{in}$-window.

## Summary and outlook

In summary, we have shown that the RIXS spectra as a function of energy loss are practically invariant to their integration over the $h\nu_{in}$-window equal to the core-hole lifetime. This fact permits the new generation of the RIXS instrumentation based on the $h\nu^2$-concept to utilize incident synchrotron radiation over the whole lifetime-wide bandwidth which is much larger than the required energy-loss resolution, thereby breaking the photon-count limit of the conventional RIXS instrumentation. At present the $h\nu^2$-concept is seen as the optimal way to keep RIXS going towards progressively refining energy scale of charge-neutral excitations without any compromise on the precision and fullness of the spectroscopic information. A practical scheme of an imaging-$h\nu^2$ spectrometer tailored to modern diffraction-limited synchrotron sources will be published in our follow-up paper.

## Acknowledgements

We thank Abhishek Nag for the support on the RIXS measurements, and Thorsten Schmitt for the fruitful discussions.

# References

1. "Magnetic excitations and phase separation in the underdoped La2-xSrxCuO4 superconductor measured b resonant inelastic X-ray scattering", L. Braicovich, J. van den Brink, V. Bisogni, M. Moretti Sala, L.J.P. Ament, N. B. Brookes, G. M. De Luca, M. Salluzzo, T. Schmitt, V. N. Strocov and G. Ghiringhelli. Phys. Rev. Lett. 04 (2010) 077002

2. "Intense paramagnon excitations in a large family of high-temperature superconductors", . Le Tacon, G. Ghiringhelli, J. Chaloupka, M. Morretti Sala, V. Hinkov, M. W. Haverkort, M. Minola, M. Bakr, K. J. Zhou, S. Blanco-Canosa, C. Monney, Y. T. Song, G. L. Sun, C. T. Lin, G. M. De Luca, M. Salluzzo, G. Khaliullin, T. Schmitt, L. Braicovich and B. Keimer. Nat. Phys. 7 (2011) 725

3. "Spin-orbital separation in the quasi-one-dimensional Mott insulator Sr2CuO3", J. Schlappa, K. Wohlfeld, K. J. Zhou, M. Mourigal, M. W. Haverkort, V. N. Strocov, L. Hozoi, C. Monney, S. Nishimoto, S. Singh, A. Revcolevschi, J. -S. Caux, L. Patthey, H. M. Ronnow, J. van den Brink and T. Schmitt. Nature 485 (2012) 485

4. "Long-range incommensurate charge fluctuations in (Y,Nd)Ba2Cu3O6+x", G. Ghiringhelli, M. Le Tacon, M. Minola, S. Blanco-Canosa, C. Mazzoli, N. B. Brookes, G. M. De Luca, A. Frano, D. G. Hawthorn, F. He, T. Loew, M. Moretti Sala, D. C. Peets, M. Salluzzo, E. Schierle, R. Sutarto, G. A. Sawatzky, E. Weschke, B. Keimer and L. Braicovich. Science 337 (2012) 821

5. "Dispersive charge density wave excitations in Bi2Sr2CaCu2O8+x", L. Chaix, G. Ghiringhelli, Y. Y. Peng, M. Hashimoto, B. Moritz, K. Kummer, N. B. Brookes, Y. He, S. Chen, S. Ishida, Y. Yoshida, H. Eisaki, M. Salluzzo, L. Braicovich, Z. -X. Shen, T. P. Devereaux and Wei-Sheng Lee. Nat. Phys. 13 (2017) 952

6. "Three-dimensional collective charge excitations in electron-doped copper oxide superconductors", M. Hepting, L. Chaix, E. W. Huang, R. Fumagalli, Y. Y. Peng, B. Moritz, K. Kummer, N. B. Brookes, W. C. Lee, M. Hashimoto, T. Sarkar, J.-F. He, C. R. Rotundu, Y. S. Lee, R. L. Greene, L. Braicovich, G. Ghiringhelli, Z. -X. Shen, T. P. Devereaux and Wei-Sheng Lee. Nature 563 (2018) 374

7. "Dynamical charge density fluctuations pervading the phase diagram of a Cu-based high-Tc superconductor", R. Arpaia, S. Caprara, R. Fumagalli, G. De Vecchi, Y. Y. Peng, E. Andersson, D. Betto, G. m. De Luca, N. B. Brookes, F. Lombardi, M. Salluzzo, L. Braicovich, C. Di Castro, M. Grilli and G. Ghiringhelli. Science 365 (2019) 906

8. "Concept of a Spectrometer for Resonant Inelastic X-ray Scattering with Parallel Detection in Incoming and Outgoing Photon Energies", V. N. Strocov, J. Synchrotron Rad. 17 (2010) 103

9. "A multiplexed high-resolution imaging spectrometer for resonant inelastic soft X-ray scattering spectroscopy" Tony Warwick, Yi-De Chuang, Dmitriy L. Voronov and Howard A. Padmore. J. Synchrotron Rad. (2014). 21, 736–743

10. "Multiplexed high resolution soft x-ray RIXS" Y.-D. Chuang, C. Anderson, M. Benk, K. Goldberg, D. Voronov, T. Warwick, V. Yashchuk, and H. A. Padmore. AIP Conference Proceedings **1741** (2016) 050011

11. The I21-RIXS Beamline at Diamond Light Source, UK: https://www.diamond.ac.uk/Instruments/Magnetic-Materials/I21.html

12. "Spin-orbit excitations in CoO" P. M. Sarte, M. Songvilay, E. Pachoud, R. A. Ewings, C. D. Frost, D. Prabhakaran, K. H. Hong, A. J. Browne, Z. Yamani, J. P. Attfield, E. E. Rodriguez, S. D. Wilson, and C. Stock. Phys. Rev. B 100 (2019) 075143

13. "Natural widths of atomic $K$ and $L$ levels, $K_\alpha$ X-ray lines and several $KLL$ Auger lines" M. O. Krause and J. H. Oliver. J. Phys. Chem. Ref. Data 8 (1979) 329

14. "Highly Efficient Beamline and Spectrometer for Inelastic Soft X-ray Scattering at High Resolution" C. H. Lai, H. S. Fung, W. B. Wu, H. Y. Huang, H. W. Fu, S. W. Lin, S. W. Huang, C. C. Chiu, D. J. Wang, L. J. Huang, T. C. Tseng, S. C. Chung, C. T. Chen, D. J. Huang. J. Synchrotron Rad. **21** (2014) 325.

15. "SAXES, a high resolution spectrometer for resonant x-ray emission in the 400–1600eV energy range" G. Ghiringhelli, A. Piazzalunga, C. Dallera, G. Trezzi, L. Braicovich, T. Schmitt, V. N. Strocov, R. Betemps and L. Patthey. Rev. Sci. Instr. **77** (2006) 113108

16. "Numerical optimization of spherical variable-line-spacing grating X-ray spectrometers".V. N. Strocov, T. Schmitt, U. Flechsig, L. Patthey and G. S. Chiuzbăian. J. of Synchrotron Rad. **18** (2011) 134